\documentclass[aps,prd,eqsecnum,showpacs,amsmath]{revtex4}
\usepackage[dvips]{color,graphicx,lscape}
\usepackage{amsfonts,amssymb,theorem}
\makeatletter

\@addtoreset{equation}{section}
\makeatletter

\newcommand{\qed}{\hbox{\rule[-2pt]{6pt}{6pt}}}
\newcommand{\D}{{\rm d}}

{\theorembodyfont{\upshape}
\newtheorem{Prop}{Proposition}}
{\theorembodyfont{\upshape}
}
{\theorembodyfont{\upshape}
\newtheorem{lm}{Lemma}}
{\theorembodyfont{\upshape}
}
{\theorembodyfont{\upshape}
}
{\theorembodyfont{\upshape}
}

\newcommand{\dalm}{\kern1pt\vbox{\hrule height 0.9pt\hbox{\vrule width
0.9pt\hskip 2.5pt\vbox{\vskip 5.5pt}\hskip 3pt\vrule width 0.3pt}\hrule height
0.3pt}\kern1pt}

\begin{document}

\title{
Nonstaticity with type II, III, or IV matter field in $f(R_{\mu\nu\rho\sigma},g^{\mu\nu})$ gravity
}

\author{Hideki Maeda}
\email{h-maeda@hgu.jp}


\affiliation{
Department of Electronics and Information Engineering, Hokkai-Gakuen University, Sapporo 062-8605, Japan. 
}

\date{\today}

\begin{abstract}
In all $n(\ge 3)$-dimensional gravitation theories whose Lagrangians are functions of the Riemann tensor and metric, we show that static solutions are absent unless the total energy-momentum tensor for matter fields is of type I in the Hawking-Ellis classification.
In other words, there is no hypersurface-orthogonal timelike Killing vector in a spacetime region with an energy-momentum tensor of type II, III, or IV.
This asserts that, if back-reaction is taken into account to give a self-consistent solution, ultra-dense regions with a semiclassical type-IV matter field cannot be static even with higher-curvature correction terms.
As a consequence, a static Planck-mass relic is possible as a final state of an evaporating black hole only if the semiclassical total energy-momentum tensor is of type I.
\end{abstract}



\maketitle



In order to understand what happens in extremely dense and curved spacetime regions such as the final stage of gravitational collapse or near the Big-Bang, descriptions of spacetimes in quantum gravity are indispensable.
However, a complete quantum theory of gravity is still not in our hands.
A possible approach in these circumstances is a semi-classical method taking quantum effects into the classical field equations.
A gravitational action describing a semi-classical spacetime must contain higher-curvature correction terms.
Then, one may consider an energy-momentum tensor of a semi-classical matter field in the resulting modified gravitational field equations.

Since observational and experimental results are still poor in such an extremely high-energy regime, a variety of modified theories of gravity and models of semi-classical matter fields have been proposed until now.
In such a rather chaotic situation, piling up solid mathematical results that hold in a wide class of gravitation theories without specifying a matter field is an important and secure activity as a physics research.

Hawking and Ellis classified an energy-momentum tensor $T_{\mu\nu}$ into four types (type I--IV) depending on the properties of its eigenvectors in a four-dimensional spacetime~\cite{Hawking:1973uf}.
Remarkably, this classification into four types is also valid in arbitrary $n(\ge 3)$ dimensions~\cite{srt1995,rst2004}. 
Properties of the eigenvectors of type-I--IV energy-momentum tensors are summarized in Table~\ref{table:scalar+1}. 
\begin{table}[htb]
\begin{center}
\caption{\label{table:scalar+1} Eigenvectors of type-I--IV energy-momentum tensors. (See appendix in~\cite{Maeda:2018hqu} for more details.)}
\begin{tabular}{|c|c|c|}
\hline
Type & Eigenvectors \\ \hline\hline
I & 1 timelike, $n-1$ spacelike \\ \hline
II & 1 null (doubly degenerated), $n-2$ spacelike \\ \hline
III & 1 null (triply degenerated), $n-3$ spacelike \\ \hline
IV & 2 complex, $n-2$ spacelike \\ 
\hline
\end{tabular} 
\end{center}
\end{table} 

Among all the Hawking-Ellis types, type-III and type-IV matter fields inevitably violate all the standard energy conditions~\cite{Maeda:2018hqu}.
While a perfect fluid is a well-known type-I matter field, a Maxwell field and a minimally coupled scalar field can be type I and II~\cite{Stephani:2003,srt1993}.
Also, a null dust is a typical type-II matter field and its generalization to include a spin of the null source, namely a gyraton~\cite{Bonnor:1970b}, is type III~\cite{psm2018,mmv2019}.
In contrast, any classical type-IV matter field is not known.
However, renormalized expectation values of the energy-momentum tensor are often of type IV~\cite{roman1986,m-mv2013}.

In this letter, we will prove non-existence of static solutions with a matter field of type II, III, or IV in a wide class of gravitation theories in arbitrary $n(\ge 3)$ dimensions.
Our conventions for curvature tensors are $[\nabla _\rho ,\nabla_\sigma]V^\mu ={R^\mu }_{\nu\rho\sigma}V^\nu$ and $R_{\mu \nu }={R^\rho }_{\mu \rho \nu }$.
The signature of the Minkowski spacetime is $(-,+,\ldots,+)$, and Greek indices run over all spacetime indices.
We adopt units such that $c=8\pi G=1$.

A static spacetime is defined by the existence of a hypersurface-orthogonal timelike Killing vector $\xi^\mu$ and its metric can be written as 
\begin{eqnarray}
g_{\mu\nu}\D x^\mu \D x^\nu=-\Omega(y)^{-2}\D t^2+g_{ij}(y)\D y^i\D y^j, \label{eq:structure}
\end{eqnarray}
where $y^i=x^i~(i=1,2,\cdots,n-1)$ are spacelike coordinates, $t=x^0$, and $\xi^\mu(\partial/\partial x^\mu)=\partial/\partial t$.
First we show the following lemma for later use.
\begin{lm}
\label{Prop:nonstaticity}
Consider a gravitation theory with the field equations ${\cal E}_{\mu\nu}=T_{\mu\nu}$ in $n(\ge 3)$ dimensions.
If ${\cal E}_{0i}=0$ holds for the spacetime (\ref{eq:structure}), there is no solution with an energy-momentum tensor $T_{\mu\nu}$ of type II, III, or IV.
\end{lm}
{\it Proof}. 
${\cal E}_{0i}=0$ gives $T_{0i}=0$ by the field equations, which shows that a timelike Killing vector $\xi^\mu(\partial/\partial x^\mu)=\partial/\partial t$ is an eigenvector of $T_{\mu\nu}$.
As seen in Table~\ref{table:scalar+1}, only type~I admits a timelike eigenvector among all the Hawking-Ellis types.
\qed

\bigskip

In this letter, we focus on a class of gravitation theories described by the following action:
\begin{align}
S=\frac12\int \D ^nx\sqrt{ -g} f(R_{\mu\nu\rho\sigma},g^{\mu\nu})+S_{\rm m},\label{action}
\end{align}
where $f$ is an arbitrary function of the Riemann tensor $R_{\mu\nu\rho\sigma}$ and metric $g^{\mu\nu}$.
By the variational principle, the action~(\ref{action}) gives the field equations ${\cal E}_{\mu\nu}=T_{\mu\nu}$~\cite{footnote1}, where $T_{\mu\nu}$ is obtained from the matter action $S_{\rm m}$ and 
\begin{align}
&{\cal E}^{\mu\nu}=R^{(\mu}_{~~\lambda\rho\sigma}{\cal {\bar F}}^{\nu)\lambda\rho\sigma}-2\nabla_\rho\nabla_\sigma {\cal {\bar F}}^{\rho(\mu\nu)\sigma}-\frac12fg^{\mu\nu}. \label{def-E}
\end{align}
Here ${\cal {\bar F}}^{\mu\nu\rho\sigma}$ is defined by 
\begin{align}
{\cal {\bar F}}^{\mu\nu\rho\sigma}:=&\frac12({\cal F}^{[\mu\nu][\rho\sigma]}+{\cal F}^{[\rho\sigma][\mu\nu]}),\label{def-barF} 
\end{align}
where 
\begin{align}
{\cal F}^{\mu\nu\rho\sigma}:=&\frac{\partial f}{\partial R_{\mu\nu\rho\sigma}}.\label{def-F}
\end{align}

Now we present our main result.
\begin{Prop}
\label{Prop:nonstaticity2}
There is no static solution in a gravitation theory (\ref{action}) if the energy-momentum tensor $T_{\mu\nu}$ is of type II, III, or IV.
\end{Prop}
{\it Proof}. 
Since independent non-zero components of the Riemann tensor in the spacetime (\ref{eq:structure}) are ${R}_{ijkl}$ and ${R}_{0i0j}$, independent non-zero components of ${\cal {\bar F}}^{\nu\lambda\rho\sigma}$ are ${\cal {\bar F}}^{0i0j}$ and ${\cal {\bar F}}^{ijkl}$ by Eqs.~(\ref{def-barF}) and (\ref{def-F}), which shows $R^{(0}_{~~\lambda\rho\sigma}{\cal {\bar F}}^{i)\lambda\rho\sigma}=0$.
On the other hand, since independent non-zero components of the Christoffel symbol in the spacetime (\ref{eq:structure}) are ${\Gamma}{}^0_{0i}$, ${\Gamma}{}^i_{00}$, and ${\Gamma}{}^i_{jk}$ and ${\cal {\bar F}}^{\nu\lambda\rho\sigma}$ depends only on $y^i$, we obtain
\begin{align}
\nabla_\sigma {\cal {\bar F}}^{000\sigma}=&\nabla_\sigma {\cal {\bar F}}^{00 i\sigma}=\nabla_\sigma {\cal {\bar F}}^{j 0 i\sigma}=\nabla_\sigma {\cal {\bar F}}^{j i 0\sigma}=\nabla_\sigma {\cal {\bar F}}^{0 i j\sigma}=0,\\
\nabla_\sigma {\cal {\bar F}}^{i 00\sigma}=&\partial_k {\cal {\bar F}}^{i 00k}+\Gamma^i_{kl}{\cal {\bar F}}^{l 00 k}+(2\Gamma^0_{0k}+\Gamma^l_{lk}){\cal {\bar F}}^{i00k},\\
\nabla_\sigma {\cal {\bar F}}^{\rho 0 i\sigma}=&\Gamma^\rho_{0k}{\cal {\bar F}}^{k 0 i0}+\Gamma^0_{0k}{\cal {\bar F}}^{\rho k i 0},\\
\nabla_\sigma {\cal {\bar F}}^{0 i 0\sigma}=&\partial_k {\cal {\bar F}}^{0 i0k}+\Gamma^i_{kl}{\cal {\bar F}}^{0 l 0k}+(2\Gamma^0_{0k}+\Gamma^l_{lk}){\cal {\bar F}}^{0 i0k},\\
\nabla_\sigma {\cal {\bar F}}^{k i j\sigma}=&\partial_l {\cal {\bar F}}^{k i jl}+(\Gamma^k_{00}{\cal {\bar F}}^{0 i j0}+\Gamma^k_{lm}{\cal {\bar F}}^{m i jl})+(\Gamma^i_{00}{\cal {\bar F}}^{k 0 j0}+\Gamma^i_{lm}{\cal {\bar F}}^{k m jl})+(\Gamma^0_{0 l}+\Gamma^m_{m l}){\cal {\bar F}}^{k i jl}
\end{align}
and 
\begin{align}
\nabla_\alpha {\cal {\bar F}}^{00i\sigma}=&\nabla_0 {\cal {\bar F}}^{k 0i0}=\nabla_l {\cal {\bar F}}^{k 0im}=0, \\
\nabla_\alpha {\cal {\bar F}}^{k 0i\sigma}=&\partial_\alpha {\cal {\bar F}}^{k 0i\sigma}+\Gamma^k_{\alpha l}{\cal {\bar F}}^{l 0i\sigma}+\Gamma^0_{\alpha\beta}{\cal {\bar F}}^{k\beta i \sigma}+\Gamma^i_{\alpha\beta}{\cal {\bar F}}^{k 0 \beta\sigma}+\Gamma^\sigma_{\alpha\beta}{\cal {\bar F}}^{k 0i\beta},\\
\nabla_\alpha {\cal {\bar F}}^{k 0i0}=&\partial_\alpha {\cal {\bar F}}^{k 0i0}+\Gamma^k_{\alpha l}{\cal {\bar F}}^{l 0i0}+\Gamma^0_{\alpha 0}{\cal {\bar F}}^{k0 i 0}+\Gamma^i_{\alpha l}{\cal {\bar F}}^{k 0 l 0}+\Gamma^0_{\alpha 0}{\cal {\bar F}}^{k 0i0},
\end{align}
which depend only on $y^i$.
Using the above expressions, one can show $\nabla_\rho\nabla_\sigma {\cal {\bar F}}^{\rho 0i\sigma}=\nabla_\rho\nabla_\sigma {\cal {\bar F}}^{\rho i0\sigma}=0$ by direct calculations.
By these results and $g^{0i}=0$, ${\cal E}^{0i}={\cal E}_{0i}=0$ is concluded.
Then, the theorem follows from Lemma~\ref{Prop:nonstaticity}.
\qed

\bigskip

Proposition~\ref{Prop:nonstaticity2} is a generalization of the claim in~\cite{hall1993} in general relativity with $n=4$.
The action (\ref{action}) describes a wide class of gravitation theories containing general relativity, Lovelock gravity~\cite{lovelock}, $f(R)$ gravity~\cite{f(R)}, the most general quadratic gravity~\cite{Salvio2018}, and so on.
An example of the theories {\it not} described by the action (\ref{action}) is topologically massive gravity (TMG) in three dimensions~\cite{tmg}.
The action of TMG is given by
\begin{align}
S_{\rm TMG}=& \frac{1}{2}\int \D^3x\sqrt{-g}\biggl[R-2\Lambda+\frac{1}{2\mu}\varepsilon^{\lambda\mu\nu}\Gamma^\rho_{\lambda\sigma}\biggl(\partial_\mu\Gamma^\sigma_{\rho\nu}+\frac23 \Gamma^\sigma_{\mu\tau}\Gamma^\tau_{\nu\rho}\biggl)\biggl]+S_{\rm m}, \label{TMG3}
\end{align}
where $\Lambda$ is the cosmological constant and $\mu$ is a coupling constant.
$\varepsilon^{\lambda\mu\nu}$ is the densitized Levi-Civita tensor with $\varepsilon^{012}=1/\sqrt{- g}$.
The field equations given from the action (\ref{TMG3}) are ${\cal E}_{\mu\nu}=T_{\mu\nu}$ with 
\begin{align}
&{\cal E}_{\mu\nu}=G_{\mu\nu}+\Lambda g_{\mu\nu}+\frac{1}{\mu}C_{\mu\nu},\\
&C_{\mu\nu}:=\frac12 \varepsilon_\mu^{~\rho\sigma}\nabla_\rho\biggl(R_{\sigma\nu}-\frac14 g_{\sigma\nu}R\biggl).
\end{align}
where $C_{\mu\nu}$ is the Cotton tensor.
In TMG, the assumption ${\cal E}_{0i}=0$ in Lemma~\ref{Prop:nonstaticity} does not hold in general for the spacetime (\ref{eq:structure}) due to the term $\varepsilon_\mu^{~\rho\sigma}$.

Proposition~\ref{Prop:nonstaticity2} shows that solutions in the system (\ref{action}) must be stationary or dynamical in a region where a matter field is of type II, III, or IV.
In particular, solutions with spherical, planar, or hyperbolic symmetry must be dynamical because their metrics can be written in diagonal forms, so that stationarity reduces to staticity in such solutions.
Note, however, that static solutions are possible with multiple matter fields of type II, III, or IV if the total energy-momentum tensor becomes type I. 
A spherically symmetric static solution in general relativity with two null dusts in~\cite{gergely1998} is such an example.

The present result asserts that, if back-reaction is taken into account to give a self-consistent solution, ultra-dense semiclassical regions with a single type-IV matter field cannot be static even with higher-curvature correction terms in the action.
As a consequence, a static Planck-mass relic, which is a hypothetical configuration as a final state of an evaporating black hole~\cite{MacGibbon:1987my}, is possible only if the semiclassical total energy-momentum tensor is of type I.
As another consequence, one can safely assume that a semi-classical matter field is of type I whenever the spacetime is static, which simplifies the analysis drastically.

\subsection*{Acknowledgements}
The author thanks Yuuiti Sendouda for helpful communications.


\end{document}